# SignaApp a modern alternative to support signwriting notation for sign languages

José Rafael Rojano-Cáceres [0000-0002-3878-4571] Alejandra Rivera-Hernández [https://orcid.org/0009-0004-8739-8601]

Universidad Veracruzana, Av. Xalapa s/n, Veracruz, México
`rrojano@uv.mx, alejandra.rihe99@gmail.com`

**Abstract.** The present work reports the development of an application called Signa App, which was designed following the philosophy of User-Centered Design. Signa App aims to provide a mobile platform for editing and creating texts in SignWriting notation. The proposal was based on the lack of a mobile application that is usable for Deaf individuals who use sign language.
The application was tested with adults, children, and adolescents, and the results showed a high degree of acceptance and ease of use. The app has already been introduced to the SignWriting user community, receiving positive feedback. Likewise, the application is available on the Google Play Store.

## 1    Introduction

People who live with hearing impairment can be characterized from the medical field as having varying degrees of hearing loss. These degrees generally classified as mild, moderate, or severe. It is estimated that within this condition 1,5 billion people are affected worldwide[1].

Within this categorization of human hearing range, it is possible to identify people who, due to a severe loss of the sense of hearing from early childhood, are not able to develop oral language naturally. Because of this condition they can acquire and develop a non-oral language, as is the case of sign language. These people, who acquire sign language as their natural language, identify themselves as Deaf signers [2].

Sign languages are characterized by the use of movement and hands' configuration, use body space, gestures and sight as means of transmitting information. Because these languages are practiced by and for Deaf people, they do not use the auditory medium in their production.

Consequently, the writing of these languages might not make sense since they do not use a system based on alphabetic characters or spellings as in the case of most hegemonic languages, but in any case they could be non-linear visual systems, based on pictographic, gestural or iconic representations [3]. For example, see **Fig. 1**, which shows a drawing sketch with a person signing "help". This sign is executed or produced with the configuration of the dactylological letter "A" with the dominant hand, resting on the base hand, which stay in a supine and extended position, it should concludes with an arrow direction that indicates who the action falls on, that is, who is helped. It should be clarified that the direction of movement has not been detailed in the gestural pictogram.

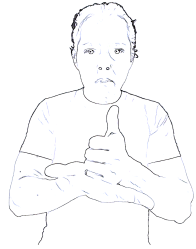

**Fig. 1.** Realistic gestural sign pictogram for "help". Source: Authors.

Therefore, it can be inferred that both oral and sign languages are naturally born in agraphs. The fact that a language has a symbolic system associated with the written production of sound or gestures is conceived as an invention subsequent to the need to transmit it through various means or any purpose required as it preservation, study, etc.. However, several linguists, as part of their formal studies, have developed notations to represent and investigate sign languages. Such is the case of the notation called HamNoSys [4] widely used in Computer Science and Linguistics, and which has antecedents in Stokoe's notation [5]. It is even possible to find rudimentary systems such as Bébian's Mimographic Notation[6].

However, these systems are far from being used or adopted in a practical way by the signing communities because they present shortcomings in their form of representation, or because of their mathematical formulation based on alphabetic symbols, particularly only useful for researchers [7]. On the contrary, currently deaf communities in more than 64 countries have opted for the use of the SignWriting notation developed by Valerie Sutton in 1974 [8], as part of a system called Sutton Movement Writing that consists of five sections [9]. This system was proposed for the reading and writing of body movement. Therefore, such notation allows for a more iconic accessible coding to signers, as Miller refers in [10]. Returning to the example of **Fig. 1A**, if you were to represent it in an iconic way, you could use the icons of a hand in supine position, the dominant hand with the dactylological configuration in "A" (thumbs up), finally an arrow towards the signer as a possible representation of the sign "help me". Next to, **Fig. 2B**, shows how it would be represented in the SignWriting notation.

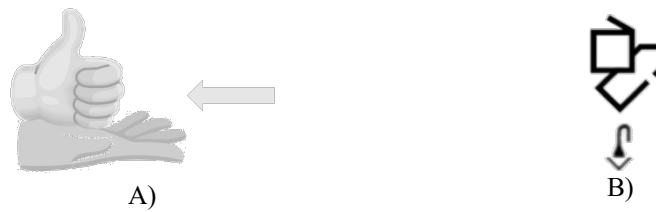

A)                                          B)

**Fig. 2.** A) Sign pictogram to help me. B) Glyph in Signwriting notation to help me

## 1.1 Problem statement

In the context of this article, although there are different developments for the afore-mentioned notations, the particular case in which we focus is that of the SignWriting notation [11] that has various software systems such as SignPuddle [12] and Sign-Maker [13]. However, regardless of the system used to transcribe sign language into a symbolic system for its written representation, the use of computer programs that are functional and usable must allow the purpose of transcription to be achieved in a simple and efficient way.

Nowadays, users of SignWriting notation use SingPuddle, which is a Web-type software designed for the creation of dictionaries and documents. However, this system, although powerful in its functionalities, has design limitations for use on mobile devices and particularly with children. This limitation has to do with the responsive aspects of the interface, which makes it difficult to use on this type of device. On the other hand, due to the altruistic nature of the project, it does not have funds, which impacts the performance of the server, making it unstable for possible use at any time.

Therefore, as part of this research, it is intended to address the problems raised, in addition to basing the proposal on User-Centered Design [14] in order to generate a usable product, which is validated by the user of mobile devices.

## 1.2 Background

According to [15], despite the popularization of SignWriting in 2015, there were no editing systems that were efficient and comprehensive enough to allow people to take notes and perform their activities satisfactorily and efficiently. Although this statement, to date, is ten years old, it is still valid, particularly if it is applied to the field of mobile platforms. A probable reason for this shortage in developments may be centered on the limited interest of researchers for the design of editors based on this notation, or the low demand of the target group to request new developments.

However, through a systematic review it is possible to identify some software proposals. First of all, there is the "official version", which is a Web platform developed by Steve Slevinski, who is a collaborator of the foundation that manages Sutton. Slevinski has been dedicated to building software for this purpose since 2004, proposing and working on the standardization of notation and various aspects of software. As a result, there is SingPuddle, which has been in force in its version 2.0 since 2017 [12]. An important part of this ecosystem is made up of SingMaker [13], which is a Web application that allows the editing and creation of new signs. In **Fig. 3**, it can be seen the view of the application on the desktop screen, and the view on a mobile device in **Fig. 4**. The loss of information is clearly visible since the interface is not responsive, among some other problems in its use such as efficiency, clarity or simplicity of its use [16].

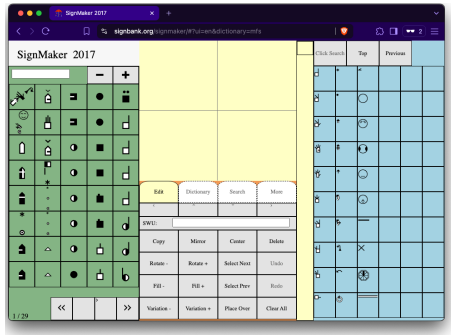

**Fig. 3.** HP monitor 1280 x 918 pixels. Source: Authors.

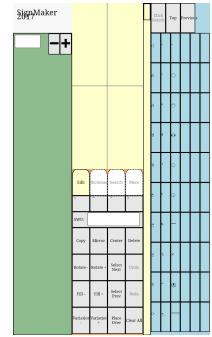

**Fig. 4.** Huawei mobile 2340x1080 pixels. Source: Authors.

Another proposal that can be found in force is JSPad [17]. This app was specifically designed for Japanese Sign Language (LSJ) writing with Sutton notation. The authors of the proposal, in addition to the typical drag-and-drop interaction system, added an alternative input method using a notation system called JJS, which is a notation system based on glosses that does not use special characters and can be written in plain text in Japanese to speed up sign writing. To date it can be find a current development for the Windows and MacOs platforms [18], in **Fig. 5**, it can be seen the general interface taken from the official site.

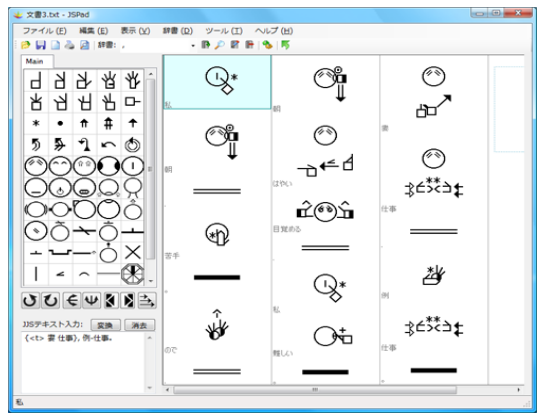

**Fig. 5.** JSPad desktop interface. Source [18].

Recently, a Web application called SignWriter was launched, which according to its author Jonathan Duncan, "is a tool for transliterating text in several languages into signwriting notation and sharing it on social networks." [19]. It is worth mentioning that the author had previously proposed the software called SignWriting Studio, which to date is no longer available [20]. In **Fig. 6** it can be seen the SignWriter interface, which presents a responsive behavior, although its function is not to create signs, but text transliteration.

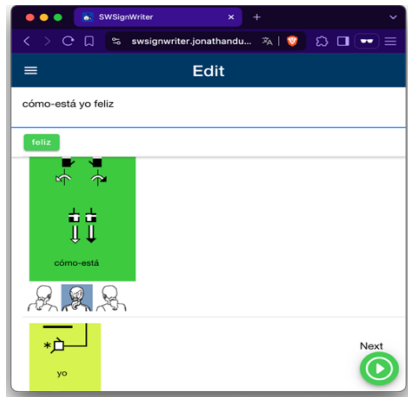

**Fig. 6.** SignWriter web interface. Source: Authors.

Finally, although in the search of the related literature it is possible to find references to other software proposals, they are no longer available for analysis. **Table 1**, presents a concentrate of the recovered applications, with their characteristics and current status, considering as a criterion for their selection that the application allows the creation of signs.

**Table 1.** Relation of software for writing signs with Sutton's notation.

| Software | | Function | Platform | Language | State | Country |
|---|---|---|---|---|---|---|
| SingPuddle | [12] | Sign Writing | Web | Multi-lingual | Active | EU |
| SingMaker | [13] | Sign Up | Web | Multi-lingual | Active | EU |
| JSPad | [18] | Sign Writing | Desk | Japanese | Active | Japan |
| SignWriter | [19] | Transliteration | Web | Unidentified | Active | Canada |
| SignWriting Studio | [20] | Sign Writing | Desk | Unidentified | Discontinued | Canada |
| SW-Edit | [21] | Sign Writing | Desk | Unidentified | Discontinued | Brazil |
| SignWriter | [22] | Sign Writing | MsDOS | Multi-lingual | Active | EU |
| Signa App | [23] | Sign Writing | Mobile | Multi-lingual | Active | Mexico |

### 1.3  Justification

Derived from the literature review, it has been found that there are several proposals for the use of software that supports sign writing in Sutton's notation; however, after analyzing and testing the software, it has been found that there are several problems in its usability, which could particularly affect deaf users. In the work of Iatskiu et al. [24], a similar study is carried out in relation to existing software for writing signs, among the results it is suggested that these tools should follow the principle of usability and HCI.

In this way, having carried out the search for applications that fulfill the function of creating, editing and modifying signs in the SignWriting notation, it was found that there is an area of opportunity for the development of an application for the mobile platform, since the existing applications, on the one hand, are not suitable to provide a good user experience due to lack of responsive management of the interface, either they have support for a particular language, or they are only available in a specific environment such as web or desktop.

It is worth mentioning that on the official SignWriting software site, it is mentioned that versions for mobile devices are in the process of development, however, to our best knowledge, there is no current version developed for this platform which allows the writing of signs from mobile devices and that provides a satisfactory user experience.

## 1.4 Objective

In this way, our main objective is to develop a mobile application that allows the creation, edition and writing of signs providing a good user experience on mobile devices, following a User-Centered Design approach.

## 1.5 Methodology

Based on the problems raised, it was selected to apply the philosophy of User-Centered Design (UCD) [14] in order to involve the user in the development process and to be able to generate a solution that was usable. As mentioned in [25], Norman, the father of user experience (UX), was the first to coin the term UCD. This is defined by four stages: a) Understanding of the user context, b) Specification of the user's requirements, c) Design of solutions and d) Evaluation. In this process, iteration is a key aspect that allows you to achieve the proposed objective when designing a product. **Fig. 7** shows the stages and their interactions.

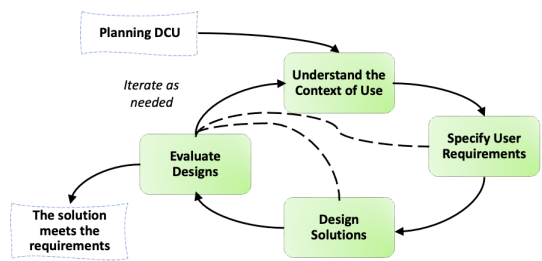

**Fig. 7.** Phases User-Centered Design. Source: Adapted from [25].

In this sense, the work that was carried out is briefly described below:

- Understanding of the user context: As a starting point, there was the participation of a Deaf person who is a user of sign language, but who has no knowledge of the notation. With the person, the analysis of their needs was carried out and a study was carried out regarding the usability of SignMaker, since this application was considered with the appropriate one to design in a mobile version. Within the con-text, an introduction to the notation was made so that the Deaf user understood what it consisted of and from there the needs were raised when interacting with the notation.

- Specification of user requirements: Derived from the previous process, we worked with the user Sordo as well as with an expert in the area of inclusive technologies. From this, aspects such as: menus, initial iconography, distribution of interaction areas and number of screens that could be required to carry out the task of editing notes with the application using SignWriting notation were defined.

- Design of solutions: For this step, prototypes were initially made on paper, these were agreed upon in several meetings with the aforementioned participants. The resulting design that was considered stable is shown in **Fig. 8**. In it there are three main screens: i) an area where the notes made will be found (main view), ii) the sign creation area, and iii) the third screen is shown when the sign was created, allowing the task of storing or exporting the sign to be carried out.

- Evaluation: In the evaluation phase, hard work was carried out with the participation of undergraduate students, Deaf users and an expert. In both cases, valuable information was obtained that allowed the successive improvement of the design by generating stable versions that were published in the Google application store.

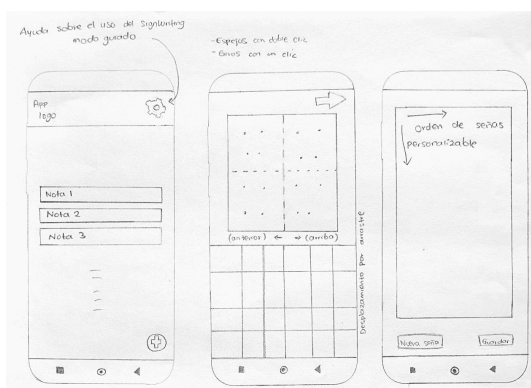

**Fig. 8.** Final prototype for the Signa App application. Source: Authors.

### 1.6    Development for Signa App

With the aim of developing a mobile application, through which it was possible to make notes using the notation created by Sutton, the implementation of the prototype on paper was carried out using the Flutter framework. Flutter[1] is a framework developed by Google for the development of cross-platform applications. Due to its multi-platform functionality, it was considered the best option since the development would first be launched on the Android platform and later it was desired to have the possibility of publishing it on IOS.

---

[1]    See: https://flutter.dev/

**Interaction flow.**
From the initial design presented in the prototype, the basic flow of three screens where the main options were had was maintained. Subsequently, it was decided to introduce a menu component that would allow adding elements that do not affect the flow of sign creation. Such is the case of language settings, tutorials, credits, among others. This process entailed the progressive adjustment of screens and menus appropriate according to the context of the application. In **Fig. 9**, is presented the screen flow map.

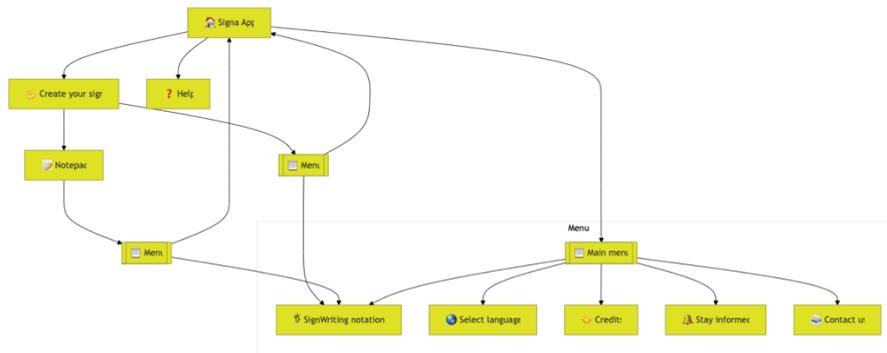

**Fig. 9.** Flow of screens in SignaApp. Source: Authors.

**Implementation.**
To ensure that the software was available to as many users as possible, the final app was published on Google's app store. It can be accessed via the https://bit.ly/signaApp link. In this sense, **Fig. 9**, shows the final result, which is still consistent with the paper prototype that was previously presented. This makes it easy to flow actions between screens. In any case, it is necessary to point out that the "Create your sign" screen is the most complex, since it contains all the symbology of Sutton's notation.

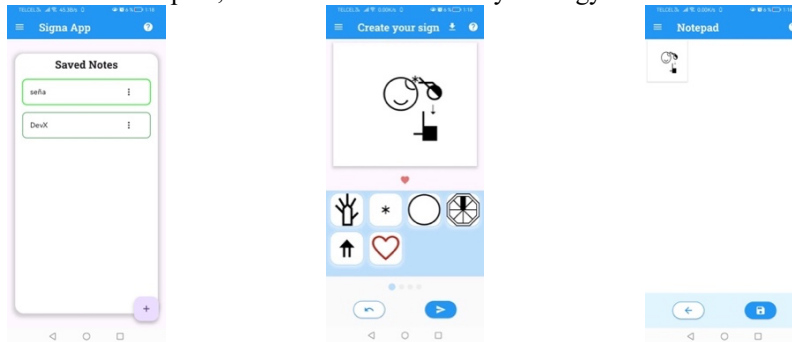

**Fig. 10.** Final screens for SignaApp. Source: Authors.

## 1.7    Evaluation

During the application development process, user evaluation was constantly maintained through various techniques. This aims to know and identify potential areas for improvement. In this regard, one of the techniques commonly used was the focus group. This technique seeks to know the perceptions, attitudes and ideas about the product. In this case, the expert was in charge of conducting the tests, which were carried out with a sample of undergraduate students in technologies. The sample had around 25 students who were Android users. Among the perceptions and opinions, the following list was obtained, and later they were corrected and published in a new version:

- It's not intuitive like toggle some buttons on, it could be put on off for when the item is selected
- When the symbol is selected, it enlarges, making it difficult to tie the pictogram with another.
- When the symbol is moved, for example by rotating it, it returns to the original shape and does not allow its position to be adjusted with respect to how the sign should be made.
- When the password is saved, it cannot be accessed, that is, it cannot be edited, it can only be renamed.
- When the image was created, it no longer allows you to download the sign.

On the other hand, considering the priority target population, that is, Deaf people, an evaluation was carried out with students from a Multiple Care Center (CAM by its acronym in Spanish). This focus group had seven students, where three of them are Deaf people and users of sign language, the rest are people living with other types of disabilities. In this regard, none of them knew the notation, so they were trained by the members of the research team. Among the tests carried out were:

- Cognitive journey: this technique made it possible to identify the ease of learning of the app's users. As a result, the reference point in terms of levels of grouping for signs had to be modified. Improved the use of controls for notation manipulation (rotation, mirroring, scaling). Likewise, an option was added for users to add their favorite signs.
- Emodiana: with this technique it was intended to measure satisfaction, however, it had to be discarded due to the cognitive complexity that its use entailed. This is because the Deaf children were not clear about the levels to identify their emotions, as well as they did not identify them clearly.
- Satisfaction questionnaire: A questionnaire was developed which contained six ad-hoc multiple-choice questions. The dimensions of it are: liking, fun, colors, videos, emotion, subjective evaluation. In this regard, the result of its application with the support of the class teacher, who is a user of sign language, is shown in Table HH. The interpretation obtained from the answers suggests that there is a high degree of general satisfaction with the application. For this conclusion, the calculation of dimensions 1 to 4 was estimated, with values ranging from 0, 0.5 and 1. In the case of the Likert scale, a value from 0 to 1 was considered, taking from 0.2 to 1. In this way the result is:

$$\text{Satisfaction} = 1+(1*50\%+.5*50\%)+1+1+(1*50\%+0.8*50\%)+(1*50\%+0.8*50\%)/6 \quad (1)$$

$$\text{Satisfaction} = 1 + 0.75 + 1 + 1 + 0.9 + 0.9 / 6 \quad (2)$$

$$\textbf{Satisfaction} = \textbf{0.92} \quad (3)$$

**Table 2.** Results of the ad-hoc satisfaction survey. The answer value for dimensions 1, 2, 3, 4 is (yes=1, no=0, more or less=0.5), the scale of values for dimensions 5 and 6 is of Likert type from the lower boundary, "I don't like it at all = 1, up to the upper boundary, I love it a lot = 5.

| Dimension | Answer | Percentage |
|---|---|---|
| 1. Like | Yes | 100% |
| 2. Fun | Yes | 50% |
| | More or less | 50% |
| 3. Color Acceptance | Yes | 100% |
| 4. Video Acceptance | Yes | 100% |
| 5. Degree of emotion | Very happy | 50% |
| | Happy | 50% |
| 6. Subjective evaluation | I love it a lot | 50% |
| | It's very good | 50% |

## 1.8 Results

It has been possible to develop a usable application, according to our evaluations, that generates a high degree of acceptance. The process of its development has been continuously validated in laboratory tests. Recently, the app was recognized by the creator of SignWriting notation, Valerie Sutton, including it on her blog as a contribution to the advancement of SignWriting. On the other hand, in Google Play statistics, it is observed which countries have downloaded the application. Mainly, Brazil is the one who leads. It has also been promoted by people in Brazil in a community of SignWriting users. However, there are also many features requests, some of which have already been attended, the most significant request includes to link the app with the official SignPuddle system, this in order to recover already created signs. Below, in **Fig. 11** it is presented the achievements.

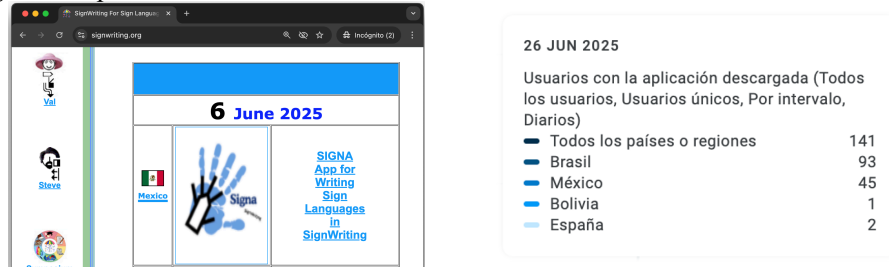

**Fig. 11.** Current achievements for Signa App. Source: Authors.

### 1.9    Conclusions

The act of communication is one of the fundamental elements that has allowed human beings to face their environment thanks to the exchange of ideas, opinions, criticisms, among others. This process was originally done through oral communications, however, when the written technology that allowed them to carry out the recording in a physical medium was created, it was possible to share and extend personal skills through the dissemination of these, both in analog and digital media. Sign language to date is a language equivalent to oral communication, therefore it has the same weakness for its distribution in the media, if it is not possible to represent it in written form. In this sense, Sutton's notation called SignWriting offers a hopeful proposal for this language to be documented, maintained and extended without depending on videos that consume a large amount of space and resources for transmission and processing.

In this sense, having technologies that allow the efficient manipulation of symbols provides the possibility for Deaf users to store and share their language in an affordable way. Thus, the development of a mobile application with support for SignWriting notation such as Signa App, allows the process to be made pleasant and efficient.

So far, the results are conclusive in terms of the possibility of use by both adult users and child users. In the case of children, it also promises to be a support for the improvement of cognitive processes, this was observed in the subsequent visits made to the CAM and witnessed by directors and teachers.

**Limitations.**

On the risk side, we identified the need to address the international community's request to link it to formal implementation. Otherwise, it will be difficult to achieve greater acceptance by its users.

**Future work.**

Future work includes continuing to continuously improve the application. Likewise, it is considered important to work on the linguistic distribution of the glyphs, since in this development the distribution of the SingMaker application was respected, but there are precedents that indicate that perhaps it should be explored more to speed up its access.

**Acknowledgments**

This work is supported by CONAHCYT through the scholarship number CVU: XXXXX. Correspondence concerning this article should be addressed to First Author.